\documentclass[review]{elsarticle}

\usepackage{lineno,hyperref}
\modulolinenumbers[5]
\usepackage[T1]{fontenc}
\usepackage{CJK}
\usepackage{xcolor}
\usepackage{amsfonts}
\usepackage{epstopdf}
\usepackage{wrapfig} 
\usepackage{subfigure}
\usepackage{graphicx}  
\usepackage{dcolumn}   
\usepackage{bm}
\usepackage{float}
\journal{Journal of \LaTeX\ Templates}
\newcommand{\be}{\begin{equation}}
\newcommand{\ee}{\end{equation}}
 \newcommand{\bea}{\begin{eqnarray}}
 \newcommand{\ena}{\end{eqnarray}}

\begin{document}

\begin{frontmatter}

\title{Thermodynamics and weak cosmic censorship conjecture in (2+1)-dimensional  regular black hole with nonlinear electrodynamics sources}

\author[mymainaddress]{Yi-Wen Han}

\author[mymainaddress]{Ming-Jian Lan}
\author[mymainaddress]{Xiao-Xiong Zeng}
\cortext[mycorrespondingauthor]{Corresponding author}
\ead{xxzengphysics@163.com}
\address[mymainaddress]{Department of physics, college of computer science and electronic information engineering, Chongqing Technology and Business University, Chongqing 400067, China }

\begin{abstract}

We study the dynamical behavior of spinor particles, and  get  the energy-momentum relation of charged particles by solving the Dirac equation. Based on the  energy-momentum relation,  we  investigate the laws of thermodynamics and the weak cosmic
censorship conjecture for the (2+1)-dimensional regular black hole with nonlinear electrodynamics sources in   the normal phase space and the extended phase space. Our results show that  the first law of thermodynamics as well as the weak cosmic censorship conjecture are valid in both the phase spaces. However, the second law of thermodynamics  is violated in the extended phase space, implying that the first law and weak cosmic censorship conjecture do not depend on the phase space while the second law depends. In addition, in the extended phase space, we find the configurations of the extremal and near-extremal black holes will not be changed for the final states and  initial states are the same.
\end{abstract}

\begin{keyword}
\texttt{Thermodynamics;  Weak cosmic censorship conjecture; Extended phase space; Regular  black hole}
\MSC[2010] 00-01\sep  99-00
\end{keyword}
\end{frontmatter}

\section{Introduction}

Based on the ideas of Henneaux et al. \cite{Henneaux2}, Caldarelli et al. \cite{Christodoulouprl} took the lead in regarding  the cosmological constant as a state parameter of the thermodynamic systems, and shown that the first law of thermodynamics for  the Kerr-Newman-AdS space time   was $dM=TdS+\Omega dJ+\Phi dQ+\Theta d\Lambda$, in which, $\Lambda$  is the generalized force conjugating to the state parameter. Since then, more and more investigations have shown that a variable cosmological  parameter   can  enrich the thermodynamic state functions \cite{Wang3,Sekiwa4, Banerjee5, Sharif6,Teitelboim, Gubser, Gibbons} so that the phase spaces of the thermodynamic system are extended, which is different from  previous treatments where the   cosmological parameter is a constant \cite{Cai,Jahani,Johnson,Caceres,A. Mandal,H. Hendi}. Recently, some authors claimed that the cosmological constant   can be used as the pressure   of the thermodynamic system \cite{D. Kastor,Dolan8}, and its thermodynamic conjugate  can be defined naturally as volume. In this framework, the mass is not the internal energy but the enthalpy. On this premise, the thermodynamics and phase transitions  of a class of black holes with cosmological constants have been  studied extensively in the extended phase space \cite{David9, Robie, David,Parthapratim,X. Mo,Zeng:2016fsb,Zeng:2016aly,Zeng:2016sei,Zeng:2015tfj}. However, most of them focused on only the first law  of thermodynamics. There was little work to check whether the second law of thermodynamics as well as weak cosmic censorship conjecture  are valid in the extended phase space.

Recently, Gwak \cite{Gwak10,Gwak:2019asi}
suggested that the laws of thermodynamics can be checked   by throwing charged particles into black holes in the framework of extended phase space.
Their results showed that the second law of thermodynamics was violated as the contributions of pressure and volume are considered. They also investigated the weak cosmic censorship conjecture  and found that
 it was valid. However, different from the case in the normal phase space, the configurations of the black hole are found to be not changed for both the extremal and near-extremal black holes.
 Subsequently, this work was extended to Born-Infeld-anti-de Sitter black hole, in which the contribution of the vacuum polarization energy was considered besides the pressure \cite{zeng}. In addition,   \cite{Gwak10}  has also been extended to the case that the absorbed particles are fermions, for which the Dirac equation was used to obtain the energy-momentum relation \cite{Y. Chen,Yun,Chen:2019pdj}. In this paper, we intend to extend the  idea in \cite{Gwak10} to the (2+1)-dimensional regular black hole with nonlinear electrodynamics sources. Unlike previous studies, this space-time is coupled by Einstein's gravity and nonlinear electrodynamics. We want to explore whether nonlinear electrodynamics affect the thermodynamics and the weak cosmic
censorship conjecture,  which  have not been reported so far. We will pay attention to the case that the   absorbed particles are fermions so that the Dirac equation will be used to investigate the dynamical behavior of the particles.

The present work is organized as follows. In section 2, the relation between the  energy and momentum of  absorbed fermions is obtained by solving the Dirac equation. In section 3, we will study thermodynamic laws and weak cosmic censorship conjecture in the normal phase space. Section 4 will extend the investigations on thermodynamic laws and weak cosmic censorship conjecture  to the extended phase space. Finally, the conclusion is presented in section 5. Throughout the paper, the units $\pi=\hbar=c=1$ are used.

\section{Energy and momentum of an absorbed  fermion }
\label{secbh}
The action of the $(2+1)$-Einstein gravity coupled with nonlinear electrodynamics is given by \cite{Fan:2016rih, Yun He, Hendi:2016pvx}
\be\label{sehy11}
S=\int d^3x\sqrt{-g}\left[\frac{R-2\Lambda}{16\pi}+L(F)\right],
\ee
where $g$ is the determinant of the metric tensor, $R$ is the Ricci scalar, $\Lambda=-1/l^2$ is the cosmological constant,  $F_{\mu \nu }=A_{\nu ,\mu }-A_{\mu ,\nu }$, and $L(F)$ is the Lagrangian of the nonlinear electrodynamics with $F=F^{\mu\nu}F_{\mu\nu}$ \cite{Cataldo}.
 From Eq.(\ref{sehy11}), we can obtain the solutions of the (2+1)-dimensional regular black hole with nonlinear electrodynamics sources, that is
\be\label{sehy12}
ds^2=-f(r)dt^2+f(r)^{-1}dr^2+r^2d\phi^2,
\ee
where
\be\label{sehy13}
f(r) = r^2/l^2 - m - q^2 \log(q^2/a^2 l^2 + r^2/l^2)
\ee
in which  $m$ and $q$ are the  parameters that relate to the mass and charge of the black hole.
The strength of electric field is
 \be\label{sehy14}
 E(r)=\frac{a^4 q r^3}{16 \pi  \left(q^2+a^2 r^2\right)^2}.
  \ee
After integration with the relation $A_t(r)=\int_{r}^{\infty} E(x)dx$ \cite{Yun He}, we can get the non-vanishing component of the vector potential
 \be\label{sehy15}
A_t(r)=-\frac{a^2 q r^2-q^3  \log\left[q^2+a^2 r^2\right]}{32 \pi }.
  \ee
By using the formula given in  \cite{Miskovic}, we can get the electric charge of the black hole, $Q=8q$. Comparing with the static, (2+1)-dimensional black hole, we also  can easily obtain the mass of the regular black holes,  $M=m/8$. Replacing $m$ and $q$ in Eq.(\ref{sehy13}) with $M$ and $Q$  respectively, the metric function can be  rewritten as
\be\label{sehy16}
F(r)=-8 M+\frac{r^2}{l^2}-\frac{1}{64} Q^2 \log\left(\frac{Q^2}{64 a^2 l^2}+\frac{r^2}{l^2}\right).
 \ee

 Now, we  investigate the  dynamical of a charged  fermion as it is swallowed by the (2+1)-dimensional regular black hole with nonlinear electrodynamics sources. The Dirac equations for electromagnetic field in curved spacetime can be expressed as
\be\label{sehy17}
i\gamma ^{\mu }\left( \partial _{\mu }+\Omega _{\mu }-\frac{i}{\hbar }%
eA_{\mu }\right)\Psi -\frac{\mu_0 }{\hbar }\Psi =0, \label{dirac}
\ee
where $u_0$ and $e$ are respectively the rest mass and charge of a fermion, $\Omega _\mu = \frac{i}{2}\Gamma _\mu ^{\rho \tau} \sum _{\rho \tau } $, $\sum _{\rho\tau } =\frac{i}{4}\left[ {\gamma ^\rho ,\gamma ^\tau } \right]$, $\gamma^\mu $ matrices satisfy $\left\{ {\gamma ^\mu ,\gamma ^\nu }\right\} = 2g^{\mu \nu}I$, in which $I$ is the unit  matrix. In order to get the solution of the Dirac equation, we have to choose    matrices $\gamma ^\mu $. Here, we set
\be\label{sehy18}
\gamma ^{\mu}=\left( -iF^{-\frac{1}{2}}\sigma ^{2},F^{\frac{1}{2}}\sigma ^{1},\frac{1}{r}\sigma ^{3}\right) ,
\ee
where  $\sigma ^\mu $ is the Pauli  matrix
\begin{equation}\label{sehyhy19}
\sigma ^{1}=\left(
\begin{array}{cc}
0 & 1 \\
1 & 0%
\end{array}%
\right) ,\sigma ^{2}=\left(
\begin{array}{cc}
0 & -i \\
i & 0%
\end{array}%
\right) ,\sigma ^{3}=\left(
\begin{array}{cc}
1 & 0 \\
0 & -1%
\end{array}%
\right).
\end{equation}%

A fermion have  spin up state and a spin down state with spin $1/2$. For the sake of simplicity, we only investigate the
spin up case for the case of spin down is similar.  We use the ansatz for the two-component spinor $\Psi $ as
\begin{equation}\label{sehy20}
\label{eq10} \Psi  = \left(
{{\begin{array}{*{20}c}
 {C\left( {t,r,\phi } \right)} \hfill \\
 {D\left( {t,r,\phi } \right)} \hfill \\
\end{array} }} \right)\exp \left( {\frac{i}{\hbar }G \left(
{t,r,\phi } \right)} \right).
\end{equation}
Substituting Eq.(\ref{eq10}) into Eq.(\ref{sehy17}), we get the following two simplified equations%
\be\label{eqhy111}
C\left( \mu_0 +\frac{1}{r}\partial _{\phi }G \right)+D%
\left[ \sqrt{F}\partial _{r}G-\left( \frac{1}{\sqrt{F%
}}\partial _{t}G-\frac{1}{\sqrt{F}}e A_t \right) \right] =0,
\end{equation}%
\be\label{eqhy112}
C\left[ \sqrt{F}\partial _{r}G+\left( \frac{1}{\sqrt{%
F}}\partial _{t}G -\frac{1}{\sqrt{F}}e A_t \right) \right] +D\left( \mu_0 -\frac{1}{r}\partial _{\phi
}G\right) =0.
\end{equation}%
The above two equations have a nontrivial solution for $C$ and $D$ if and only if the determinant of the matrix coefficients is zero. So we get
\begin{equation}\label{eqhy113}
\frac{1}{r^{2}}\left( \partial _{\phi }G
\right) ^{2}-\mu_0 ^{2}+\left( \sqrt{F}\partial _{r}G\right) ^{2}-\left( \frac{1}{\sqrt{F}}\partial _{t}G -\frac{1}{\sqrt{F}}e A_t \right) ^{2}=0.  \label{wave}
\end{equation}
There are two Killing vectors
in the (2+1)-dimensional spacetime, so
$G\left( t,r,\phi \right) $ in   Eq.(13) can be separated into
\begin{equation} \label{eghy333}
G=-\omega t+L \phi +R\left( r\right)+\lambda,
\end{equation}%
where $\omega $ and $L$ are Dirac particle's energy and angular momentum
respectively, and $\lambda$ is a complex constant. Substituting Eq.(\ref{eghy333}) into Eq.(\ref{wave}), we obtain
\begin{equation}\label{eq114}
\partial _{r}R\left( r\right) =\pm \frac{1}{F}\sqrt{\left( \omega +eA_t \right) ^{2}+F\left( \mu_0 ^{2}-\frac{L^{2}}{r^{2}}%
\right) }.
\end{equation}%
In order to  study thermodynamics, we focus on only the radial momentum of particles near the horizon, i.e. $p^r\equiv g^{rr}p_r=\partial _{r}R\left( r\right)$.  From Eq. (\ref{eq114}),   we get
\begin{equation}\label{eqh0020}
\omega= |p^r|-eA_t(r_h).
\end{equation}
As done in \cite{Gwak10,
Gwak:2018akg,Gwak:2015fsa,Christodoulou}, we choose the
positive sign in front of $|p^r|$.

\section{Thermodynamics and weak cosmic censorship conjecture of (2+1)-dimensional black hole in the normal phase space }

The electrostatic potential between the black hole horizon and the infinity  is defined by $\Phi=A_t(\infty)-A_t(r_{h})$. After integration, we can obtain
\begin{equation}
\Phi_h =-\frac{Q \left(l^2 Q^2+\left(l^2 Q^2+64 a^2 r_h^2\right) \log\left(\frac{Q^2}{64 a^2}+\frac{r_h^2}{l^2}\right)\right)}{256 \left(l^2 Q^2+64 a^2 r_h^2\right)},
\end{equation}
where $r_h$ is the event horizon of the (2+1)-dimensional regular black hole with nonlinear electrodynamics sources, which is determined by $F(r_h)=0$. According to the definition of surface gravity, the Hawking temperature  can be written as
\begin{equation}\label{eqhyw1}
T_h=\frac{r_h }{128 l^2 \pi }\left(64-\frac{64 a^2 Q^2 r_h^2}{l^2 Q^2+64 a^2 r_h^2}\right).
 \end{equation}
By using the  Bekenstein-Hawking  entropy area relation, the black hole entropy can be written as
\begin{equation}\label{eqhyw2}
S_h=\frac{1}{2}\pi r_h.
 \end{equation}
From Eq.(\ref{sehy16}), we can get the mass of  the (2+1)-dimensional regular black hole with nonlinear electrodynamics sources, which can be expressed as
\begin{equation}
M=\frac{r_h^2}{8 l^2}-\frac{Q^2}{512} {\log}\left[\frac{1}{l^2}\left(\frac{Q^2}{64 a^2}+r_h^2\right)\right].
 \end{equation}

As a charged fermion is  swallowed by the regular black holes, the variations of the  internal energy and charge of the black hole system should satisfy
\begin{equation}
\omega=dM,\quad e=dQ,
\end{equation}
where the conservations of energy and  charge   have been considered. Now, Eq.(\ref{eqh0020}) can be rewritten as
\begin{equation}\label{eqyw002}
dM=\Phi_h dQ+p^r.
\end{equation}
Similarly,  the event horizon of a black hole changes as it absorbs particles,  leading to change in $F(r)$. In the new
horizon, there is also a relation, $F(r_h+dr_h)=0$. In other words,  the change of the horizon    should satisfy
\be \label{df}
dF_h=\frac{\partial F_h}{\partial M}dM+\frac{\partial F_h}{\partial Q}dQ+\frac{\partial F_h}{\partial r_h}dr_h=0.~~~~
\ee
Inserting Eq.(\ref{eqyw002}) into Eq.(\ref{df}),  $dM$ will be deleted. Interestingly, $dQ$  will de deleted too. We can get  $dr_{h}$ directly
\be \label{dr}
dr_h=\frac{4 p^r \left(l^4 Q^2+64 a^2 l^2 r_h^2\right)}{r_h \left(\left(1-a^2\right) l^2 Q^2+64 a^2 r_h^2\right)}.
\ee
With Eqs.(\ref{eqhyw2}) and  (\ref{dr}), the variation of entropy can be expressed as
\begin{equation}\label{eqh001}
dS_h=\frac{2 \pi  p^r \left(l^4 Q^2+64 a^2 l^2 r_h^2\right)}{r_h \left(\left(1-a^2\right) l^2 Q^2+64 a^2 r_h^2\right)}.
\end{equation}
From Eqs.(\ref{eqhyw1}) and (\ref{eqh001}), we get
\be
T_h dS_h=p^r.
\ee
So, the internal energy in Eq.(\ref{eqyw002}) can be rewritten as
\begin{equation}\label{eqyw44}
dM=TdS+\Phi dQ,
\end{equation}
which is  the first law of the (2+1)-dimensional regular black hole with nonlinear electrodynamics sources in the normal phase space.

So far, we have derived the first law of thermodynamics.  Next, we are going to discuss the second law of the black hole thermodynamics. For the extreme  black holes, the temperature is  zero, which implies.
\begin{equation}\label{eqtt001}
r_h=\frac{l Q}{\sqrt{-64 a^2+a^2 Q^2}}.
\end{equation}
 The  second law  for the extremal black holes is meaningless strictly for the temperature of the system is zero.  So we are interested only the  non-extremal  black hole, where  the temperature is larger then zero.  According to the Eq.(\ref{eqhyw1}), the event horizon should satisfy $r>r_h$.
In this case, $dS_h$  is positive in  Eq.(\ref{eqh001}), which shows that the second law of thermodynamics is also valid.

Furthermore, we can investigate the weak cosmic censorship conjecture, which states that the singularity should be hidden by the event horizon of the black hole for an observer located  at infinity.
 To ensure the validity of the weak cosmic censorship conjecture, the existence of an event horizon is needed. To do this, we try to test whether an event horizon exists when a fermion is absorbed by the black hole. That is, whether there are solutions for
$F(r)$.

 For the (2+1)-dimensional regular black hole with nonlinear electrodynamics sources, there is a minimum value  for $F(r)$ with the radial coordinate $r_{min}$. For the case $F(r_{min})>0$, there is not a horizon while for the case $F(r_{min})\leq0$, there are horizons always.   At $r_{min}$,
 the following relations should be satisfied, yielding
\bea
&F|_{r=r_{min}}\equiv F_{min}=\alpha\leq 0,\nonumber\\ \label{eqhy234}
&\partial_{r}F|_{r=r_{min}}\equiv F'_{min}=0,\nonumber \\
&(\partial_{r})^2 F|_{r=r_{min}}\equiv F^{\prime\prime}_{min}>0.
\ena
For the extreme black hole, $\alpha=0$, and for the near extreme black hole, $\alpha$ is a small quantity.  The inner horizon and outer horizon are distributed at either side of
$r_{min}$.  As a fermion is swollowed into the black hole, the change in mass and charge of the black hole can be written as  $(M+dM, Q+dQ)$. The locations of minimum value and the event horizon are written as $ r_{min}+dr_{min}$, $r_{h}+dr_{h}$ correspondingly. Here, we have a transformation of $F(r)$, which is labeled as $dF_{min}$. By using condition $F'_{min}=0$, at $r_{min}+dr_{min}$, $F(r)$ can be expressed as
\bea \label{eqc1}
F|_{r=r_{min}+dr_{min}}=F_{min}+dF_{min}=\alpha+\left(\frac{\partial F_{min}}{\partial M}dM+\frac{\partial F_{min}}{\partial Q}dQ\right),
\ena
 For the extremal black hole, the horizon  is located at $r_{min}$, the  Eq.(\ref{eqyw002}) is valid at  $r_{min}$. Substituting Eq.(\ref{eqyw002})  into  Eq.(\ref{eqc1}),    $dQ$ will be deleted at the same time.  Therefore, the Eq.(\ref{eqc1}) can be reduced to
\be\label{eqc6}
F_{min}+dF_{min}=-8p^r.
\ee
 Obviously, $F(r_{min}+dr_{min})$ is smaller than $ F(r_{min})$ when a charged fermion is engulfed by the black hole. This will cause extremal black holes to become into non-extremal black holes. In other words,  the weak cosmic censorship conjecture is valid  in the normal phase space.
\section{Thermodynamics and  weak  cosmic censorship conjecture in the extended  phase space}

Now, we turn to investigate thermodynamics and  weak  cosmic censorship conjecture with a variable cosmological parameter, namely $\Lambda$ is treated as the thermodynamic pressure, and its conjugate variable is the thermodynamic volume, which can be represented as respectively
\bea\label{eq16}
P&=&\frac{1}{8 \pi l^2},\\
V_h&=&\frac{\pi  r_h^2 \left(\left(1-a^2\right) l^2 Q^2+64 a^2 r_h^2\right)}{l^2 Q^2+64 a^2 r_h^2}.\label{eq17}
 \ena
From Eqs. (\ref{eqhyw1}), (\ref{eqhyw2}), (\ref{eq16}) and (\ref{eq17}),  we obtain the Smarr formula
\be
T_h S_h=2 V_h P.
\ee
Considering that $M$ has the physical  significance of enthalpy in the extended phase space\cite{D. Kastor,Dolan8}, it has a  relation to internal energy as
\be\label{eq119}
M=U + PV_h.
\ee

As a charged fermion is swollowed into the black hole, the energy and charge are supposed to be conserved. So, the change in energy and charge of the black hole system should be equal to the energy and charge of the particle. Namely
\be
\omega=dU=d(M-PV_h),\quad e=dQ.
\ee
 The energy in Eq.(\ref{eqyw002}) changes correspondingly into
\be \label{eqhyw99}
dU=\Phi_h dQ+|p^r|.
\ee
Because of the backreaction, the absorbed fermions will change the event horizon of the black hole. But the  event horizon is always determined by  $F(r)$. In other words, in the new horizon,  there is also a relation $F(r_h+dr_{h})=F(r_h)+dF_{h}=0$,  namely
\be \label{eqhyw97}
dF_{h}=\frac{\partial F_ {h}}{\partial M}dM+\frac{\partial F_ {h}}{\partial Q}dQ+\frac{\partial F_ {h}}{\partial l}dl+\frac{\partial F_ {h}}{\partial r_ {h}}dr_ {h}=0.
\ee
With Eq.(\ref{eq119}), the Eq.(\ref{eqhyw99}) can be rewritten as as
\be \label{eqhyw98}
dM-d(PV_h)=\Phi_h dQ+|p^r|.
\ee
From Eq.(\ref{eqhyw98}), we can get $dl$, and substituting  $dl$ into Eq.(\ref{eqhyw97}), we can delete it directly. Interestingly, $dQ$, and $dM$ are also eliminated simultaneously. Based on this, there is only a  relation between $|p^r|$  and  $dr_{h}$, that  is
 \be \label{eqhyw77}
dr_ {h}=-\frac{ p^r \left(l^2 Q^2+64 a^2 r_h^2\right){}^2}{16 a^4 Q^2 r_h^3}.
\ee
On this basis, the variations of entropy and volume of the black hole can be expressed as
\be\label{eqhyw88}
dS_h=-\frac{\pi p^r \left(l^2 Q^2+64 a^2 r_h^2\right){}^2}{32 a^4 Q^2 r_h^3},
\ee

\be\label{eqh002}
dV_h=\frac{\pi   p^r  \left(\left(-1+a^2\right) l^4 Q^4-128 a^2 l^2 Q^2 r_h^2-4096 a^4 r_h^4\right)}{8 a^4 Q^2 r_h^2}.
\ee
According to above formulations, we find
\be  \label{prpv}
T_h dS_h-PdV_h=|p^r|.
\ee
The internal energy in Eq.(\ref{eqhyw99}) thus would change into
\be  \label{eqh003}
dU=\Phi_h dQ + T_h dS_h-PdV_h.
\ee
 With Eq.(\ref{eq119}), we can obtain the relation between the enthalpy and internal energy in the extended phase space, that is
\be  \label{eqhy0040}
dM=d U+P dV_h+V_h dP.
\ee
Substituting Eq.(\ref{eqh003}) into Eq.(\ref{eqhy0040}), we get
\be
dM=T_h dS_h+\Phi_h dQ+V_h dP,
\ee
which is consistent with that in Eq.(\ref{eqyw44}). That is, in the extended phase space, the first law of thermodynamics still holds when charged fermions are absorbed by the black hole.

 We are willing to discuss  the second law of thermodynamics in extended phase space. From   Eq.(\ref{eqhyw88}), we find $dS_h < 0$ . Obviously, in the extended phase space, the change of entropy of the (2+1)-dimensional regular black hole with nonlinear electrodynamics sources is negative. In other words, the entropy of the black hole unavoidably decreases, which is contrary to the second law of thermodynamics. We suspect that the reason might be the contribution of the $PV_h$  term.

Next, we also can discuss the weak cosmic censorship conjecture in the extended phase space.  Considering the backreaction,  the mass $M$, charge $Q$, and AdS radius $l$ of the black hole will change into $(M+dM, Q+dQ, l+dl)$ as a charged fermion is swallowed by the black hole.  The locations of the minimum value and the event horizon will change into $ r_{min}+dr_{min}$, $ r_{h}+dr_{h}$ correspondingly. At $r_{min}+dr_{min}$, the change of $F(r)$  is subject to
\be \label{eeqc1}
F|_{r=r_{min}+dr_{min}}=\alpha+(\frac{\partial F_{min}}{\partial M}dM+\frac{\partial F_{min}}{\partial Q}dQ+\frac{\partial F_{min}}{\partial l}dl).
\ee
By using $F'_{min}=0$ in Eq.(\ref{eeqc1}),  there is also a relation at the new lowest point, which can be expressed as
\be
\partial_{r} F|_{r=r_{min}+dr_{min}}=F'_{min}+dF'_{min}=0,
\ee
implying
\be \label{eehy2}
dF'_{min}=\frac{\partial F'_{min}}{\partial Q}dQ+\frac{\partial F'_{min}}{\partial l}dl+\frac{\partial F'_{min}}{\partial r_{min}}dr_{min}=0,
\ee
where we have used the condition $F'(r_{min})=0$ in Eq.(\ref{eqhy234}). Solving Eq.(\ref{eehy2}), we get
\bea \label{pvdl}
dl&=&\frac{ \left(\left(1-a^2\right) l^5 Q^4+64 a^2 \left(2+a^2\right) l^3 Q^2 r_{min }^2+4096 a^4 l r_{min }^4\right)}{2 r_{min } \left(\left(1-a^2\right) l^4 Q^4+128 a^2 l^2 Q^2 r_{min }^2+4096 a^4 r_{min }^4\right)}{dr}\nonumber \\
&=&\frac{-128 a^4 l^3 Q r_{min }^3}{2 r_{min } \left(\left(1-a^2\right) l^4 Q^4+128 a^2 l^2 Q^2 r_{min }^2+4096 a^4 r_{min }^4\right)} {dQ}.
\ena

For the extremal black hole, the horizon is located at $r_{min}$. That is,  Eq.(\ref{eqhyw99}) is applicable at $r_{min}$.  Substituting Eq.(\ref{eqhyw99}) into Eq.(\ref{eeqc1}),
 we obtain
 \be \label{eqhy11}
F_{min}+ dF_{min}= - 8 p^r+\frac{2  dr r_h }{l^2} \left(-1+\frac{a^2 l^4 Q^4}{\left(l^2 Q^2+64 a^2 r_h^2\right){}^2}\right).
\ee
Substituting  Eq.(\ref{prpv}) into  Eq.(\ref{eqhy11}), we find
\be \label{df1}
F_{min}+ dF_{min}= \frac{2 r_{\min } \left(\left(-1+a^2\right) l^2 Q^2-64 a^2 r_{\min }^2\right) {dr} }{l^4 Q^2+64 a^2 l^2 r_{\min }^2}.
\ee
In addition,
 according to the condition
$F'_{min}=0$, we can get $Q$, that is
\be \label{eqc33}
Q=\frac{8 a r}{\sqrt{-l^2+a^2 l^2}}.
\ee
Substituting  Eq.(\ref{eqc33}) into  Eq.(\ref{df1}), we get lastly
\be \label{df2}
F_{min}+ dF_{min}=0.
\ee
Eq.(\ref{df2}) shows that when a charged fermion is absorbed by the black hole, there is not  shift for $F_{min}$. In other words, the configurations of the black holes are not changed, the extremal black holes are still extremal black holes as fermions are absorbed.

For the near-extremal black hole,  Eq.(\ref{eqhyw99}) is not applicable at $r_{min}$ for it is valid only at $r_h$. But we can expand it near  $r_{min}$ with $r_h=r_{min}+\epsilon$. Note that $p^r$ should also be expanded for it is  a function of the horizon $r_h$. To the first order, we get
\bea \label{dmfv}
dM=\frac{A+B+E}{256 l^3 \left(l^2 Q^2+64 a^2 r_{\min }^2\right)}
+\frac{(X+Y+Z) \epsilon}{4 l^3 \left(l^2 Q^2+64 a^2 r_{\min }^2\right){}^2}+O(\epsilon)^2,
\ena
in which
\bea
&&A=l^5 \left( Q^3 {dQ}+ Q^3 \log\left[\frac{Q^2}{64 a^2}+\frac{r_{\min }^2}{l^2}\right]{dQ} \right)\nonumber,\\
&&B=-4096 a^2  l r_{\min }^3{dr}+4096 a^2  r_{\min }^4{dl}+l^2 \left(64 Q^2 r_{\min }^2{dl}-64 a^2  Q^2 r_{\min }^2{dl}\right),\nonumber\\
&&E=l^3  (-64 Q^2 r_{\min } {dr} +64 a^2Q^2 r_{\min } {dr} +64 a^2 Q \log\left[\frac{Q^2}{64 a^2}+\frac{r_{\min }^2}{l^2}\right] r_{\min }^2 {dQ} ),\nonumber\\
&&X= -128 a^4  l^3 Q r_{\min }^3{dQ} +4096 a^4  l r_{\min }^4{dr}-8192 a^4 r_{\min }^5 {dl}                    ,\nonumber\\
&&Y=   Q^4 \left( l^5{dr}-a^2 l^5{dr} -2 l^4 r_{\min } {dl}+2 a^2  l^4 r_{\min }{dl}\right)               ,\nonumber\\
&&Z=Q^2 \left(128 a^2  l^3 r_{\min }^2{dr} +64 a^4  l^3 r_{\min }^2{dr} -256 a^2 l^2 r_{\min }^3{dl}\right). \nonumber
\ena
Substituting Eq.(\ref{dmfv}) into Eq.(\ref{eehy2}), we can get $F_{min}+ dF_{min}$. Substituting $dl$ in Eq.(\ref{pvdl}) in to this equation, we find
\bea\label{dfrm3}
&&F_{min}+ dF_{min}=\alpha+ \frac{ 4 r_{\min }^2 \left(\left(1-a^2\right) l^2 Q^2+64 a^2 r_{\min }^2\right)K}{H}
+O(\epsilon)^2,
\ena
where we have defined
\bea
K&=& -\left(1-a^2\right) l^4 Q^4 {dl}+64 a^2 l^2 Q \left(a^2  l{dQ}+2  {dl} Q\right) r_{\min }^2+4096 a^4  r_{\min }^4{dl}, \nonumber\\
H&=&\left(-1+a^2\right) l^9 Q^6-192 a^2 l^7 Q^4 r_{\min }^2 \nonumber\\
&-&4096 a^4 \left(3+a^2\right) l^5 Q^2 r_{\min }^4-262144 a^6 l^3 r_{\min }^6. \nonumber
\ena
Substituting Eq.(\ref{eqc33})  into Eq.(\ref{dfrm3}), we  find lastly
\bea\label{dfrm1}
F_{min}+ dF_{min}=\alpha +
O(\epsilon)^2,
\ena
which is the same as that for the extremal black holes in Eq.(\ref{df2}) for $O(\epsilon)^2$ is the high order terms of $\epsilon$, that is, the near-extremal black holes are also stable.  So we can conclude that the  weak
cosmic censorship conjecture holds for both the extremal and near-extremal black holes in the extended phase space as fermions are adsorbed.

\section{Conclusion}

In this work, we first studied the dynamics of charged fermions being swallowed by the (2+1)-dimensional regular black holes with nonlinear electrodynamics sources. We obtained the energy-momentum relation of the spinor particles by solving the Dirac equation. With this relation,  we got the first law, and the second law of thermodynamics   in the normal phase space further. We found that the laws of thermodynamics for the regular black holes held. We also discussed  the weak cosmic censorship conjecture and found it was valid in the normal phase space for there exist horizons always stoping  the singularities to be naked. Especially, the extremal black holes will evolve into the non-extremal black holes.

We also investigated the laws of thermodynamics and  weak cosmic censorship conjecture in the extended phase space, where the cosmological parameter is regarded  as an extensive variable of the thermodynamic system. We  derived the first law of thermodynamics  by making use of the conservations of energy and charge.   The second law of thermodynamics was checked too by studying the variation of the entropy of the (2+1) dimensional black hole with nonlinear electrodynamic sources. The result showed that  the entropy   decreased unavoidably. The second law thus is violated in the extended phase space, which is different from the case in the normal phase space. We  investigated the weak cosmic censorship conjecture for the  extremal and near-extremal black holes in extended phase space and found that it was valid too. But different from that in the normal phase space, the configurations of the regular black holes will not be changed as charged fermions are absorbed for the final states are still  extremal and near-extremal black holes.

\section*{Acknowledgements}{This work is supported  by the National
Natural Science Foundation of China (Grant No. 11875095), and Basic Research Project of Science and Technology Committee of Chongqing (Grant No. cstc2018jcyjA2480).}

\end{document}